# Blockchain Application on the Internet of Vehicles (IoV)

Nyothiri Aung
University College Dublin
Dublin, Ireland
nyothiri.aung@ucd.ie

Tahar Kechadi
University College Dublin
Dublin, Ireland

Tao Zhu
University of South China
Henyang, China

Saber Zerdoumi
United Nation International Solar
Energy Technology Transferring
Center

Tahar Guerbouz
University of Ouargla
Ouargla, Algeria

Sahraoui Dhelim *
University College Dublin
Dublin, Ireland
sahraoui.dhelim@ucd.ie

*Abstract*—With the rapid development of the Internet of Things (IoT) and its potential integration with the traditional Vehicular Ad-Hoc Networks (VANETs), we have witnessed the emergence of the Internet of Vehicles (IoV), which promises to seamlessly integrate into smart transportation systems. However, the key characteristics of IoV, such as high-speed mobility and frequent disconnections make it difficult to manage its security and privacy. The Blockchain, as a distributed tamper-resistant ledge, has been proposed as an innovative solution that guarantees privacy-preserving yet secure schemes. In this paper, we review recent literature on the application of blockchain to IoV, in particular, and intelligent transportation systems in general.

## I. INTRODUCTION

The Internet of Vehicles (IoV) has emerged as a result of the convergence of the Internet of Things (IoT) and Vehicular Ad-Hoc Networks (VANETs). IoV offers ubiquitous access to information to drivers and passengers while on the move [1]. IoV has many promising applications such as intelligent traffic optimization [2], sensor-based accident prediction [3], entertainment content delivery [4], and electric vehicle charging scheduling [5]. However, with the fast-growing number of connected vehicles, IoV requirements are difficult to satisfy, e.g. secure, scalable, robust, seamless information exchange among vehicles, users, and roadside infrastructures. Blockchain [6], a distributed and immutable tamper-resistant ledger, manages persistence records of data at different nodes and has the potential to deal with the data security and privacy challenges in IoV networks. As shown in Figure 1, the application of Blockchain to IoV can be roughly divided into four classes, namely security, privacy, trust computing and incentive mechanisms. For security, Blockchain can be leveraged to solve the problems of traditional vehicular network security issues, such as public key infrastructure (PKI)-based solutions and ID-based solutions. Similarly, Blockchain-enabled privacy schemes for IoV ensure the identification of the privacy of the participating vehicles and their locations. Blockchain can also serve as a distributed reputation system that stores vehicles' trust scores or store vehicle generated content. Finally, Blockchain can also be used to manage traffic-related cryptocurrencies for rewarding vehicles that helped the system. In this paper, we review recent literature on blockchain applications in IoV and intelligent transportation systems in general.

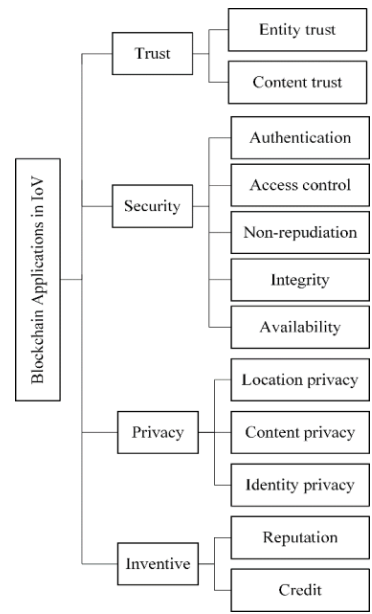

Fig. 1. Blockchain applications in IoV

## II. BLOCKCHAIN FOR IOV TRUST MANAGEMENT

Trust management refers to the methods that estimate the trustworthiness of other entities in the system. Based on the likelihood of trustworthiness assigned to a given entity, the system decides whether to interact with that arbitrary entity or not. The concept of trust is paramount in systems security, and it is discussed extensively in the network security literature [7]. In the context of vehicular networks, when the vehicle receives a message, the trust system must check the legitimacy and trustworthiness of the sender (node trust), as well as the trustworthiness of the received data or information (content trust). Considering the nature of vehicular networks, trust management plays a vital role in preventing network attacks that may cause serious consequences. For instance, in a safety related VANET application of emergency warning, the application decides to operate the emergency brakes in case of a nearby accident or road obstacle. When the emergency application receives a warning message from other nodes, it is up to the trust management system to evaluate the trust level of information, as well as the legitimacy and trustworthiness of the sender. In such a delay-sensitive application, the trust management system should immediately decide whether the warning message is coming from a legitimate node and the information is correct, or from a malicious vehicle for selfish purposes. Following the concluded decision, the trust management system announces the decision to the emergency application, the latter either operates the brake if the message was trusted or ignores the message and reports the sender to the trust authority in case of fake content. Since the conventional security mechanisms, such as cryptography, access control and authentication, cannot be used to detect dishonest nodes or fake messages, a trust management system is necessary to guarantee the security of the network and it is supposed to complement the other security measures of vehicular networks.

There are many trust management methods that guarantee the trustworthiness of network nodes in MANET [8]. However, due to the special characteristics of VANET, the MANET trust management schemes cannot be adopted for VANET, which need more efficient and scalable methods. Due to the high mobility of the VANET nodes, and the frequent changing of the network topology, it is difficult to maintain the same set of neighbours for a long time. This requires the trust management system to estimate the trust level of a giving node immediately after their short encounter, rather than maintaining a trust index of long-lasting cluster neighbours and updating the trust index, like trust management in MANET. Vehicular nodes are more likely to meet with other nodes that they have never encountered before, and cannot wait until they meet these nodes again to update their trust level, therefore trust bootstrapping is crucial in VANET [9]. Vehicular nodes need to cooperate to maintain a global trust management system, and depend on other sources, such as social networks, to estimate the trust level.

Blockchain-based trust management system leverages blockchain properties to tackle the challenges of trust management in vehicular networks. The blockchain decentralization property ensures that all the vehicular nodes participate in the computation and storage of the global trust records. This eliminates the need for a central trust authority that stores and updates trust records. The availability property ensures that the trust management system could still work even in the presence of failure of any vehicle. The immutability property guarantees that stored trust levels are not changeable unless the vehicular community agrees to update some of them by adding new transactions, hence malicious vehicles cannot tamper with their trust value on the blockchain. The transparency property gives the chance to any vehicle the ability to fetch any trust record if needed. Zhang et al. [10] introduced a blockchain-based trust management framework for IoV, the proposed system uses a reputation-based rating scheme to evaluate the trustworthiness of the vehicles and RSU. The vehicles rate the correctness of the received message and upload their rating to the RSU community, the latter computes the reputation value of the nodes who declare this event and update their reputation values in the blockchain. In case of a false message, the system decreases the reputation value of the malicious node, which weakens its influence in future decisions. A PoW and PoS hybrid consensus algorithm is used to guarantee that the high-reputed vehicles have the priority to update the blockchain, and smart contracts are used to ensure that the RSU computes the reputation scores according to the transparent rules. It is worth noting that the blockchain is stored only on the RSU, not on the vehicles. Similarly, Yang et al [11] proposed a blockchain-based distributed trust management system that leverages Bayesian Inference Model to evaluate the upcoming messages from the neighbouring nodes. The RSU community aggregates the ratings from vehicles and collaboratively compute the trust value offsets; these trust values are grouped into blocks and stored in the blockchain. This also uses a hybrid PoW and PoS consensus algorithm, where the more total value of the trust offsets (PoS), the easier RSU can compute the nonce of the hash function (PoW).

In the same vein, Liu et al [12] used a conditional privacy-preserving announcement method to propose a blockchain-based trust management system, called (BTCPS). BTCPS leverages identity-based group signature schemes to ensure conditional privacy, in case of fake message dissemination, the trusted authority can unveil the identity of the malicious vehicle through anonymous announcements of the designated public address. RSUs cooperate to maintain a distributed ledger that contains the trust value of each vehicle. When an event is detected, the source vehicle calls nearby nodes as witnesses to confirm the event. In case of approval, the witnesses reply with a response message that contains the signatures of these vehicles, the initiator sends the aggregated signatures package to the nearest RSU for verification. The RSU uses logistic regression to compute the trust values of the initiator and verify the message. After updating the trust value of the initiator (increase it in case of a true message or decrease it in case of a fake message), the elected minor RSU packs the new values into a block and tries to add the new block by providing the block ID, RSU ID, block generation timestamp, previous block's hash and the nonce. Some authorized RSUs are responsible to perform the consensus using a hybrid PoW and PBFT algorithm. Wang et al [13] proposed a blockchain-based trustworthiness computing system named (B-TSCA), where the RSUs validate the trustworthiness of the vehicles through consensus, B-TSCA uses Merkle hash tree (MHT) to implement real-time audit of vehicle attributes, such as trust level. Javaid et al [14] presented a blockchain-based trust management system that uses a physical unclonable function (PUF) to assign unique identifiers to each vehicular node and smart contract along with a dynamic proof-of-work (dPoW) consensus algorithm to store the nodes' trust scores in the blockchain. Kouicem el. [15] introduced BC-Trust, a blockchain-based trust management system for highly mobile devices, where the network nodes disseminate trust ratings about service providers to the blockchain. Hence, all the nodes can access trust information about any service provider. Singh et al [16] presented a blockchain-based distributed trust management system using smart contracts. They leverage the concept of blockchain sharding to decrease the computing overhead on the blockchain and increase the transaction throughput. The proposed system uses an incentive mechanism to encourage vehicles to report fake messages, the reward points could be spent to benefit from vehicular services, such as insurance and maintenance. Dhelim et al. [17]-[23] discussed trust computing in the context of IoP and its applications in recommendation systems and user interest mining. Similarly, Ning et al. [24]–[28] studied trust computing in cyber-physical mapping systems.

TABLE I. BLOCKCHAIN FOR TRUST MANAGEMENT

| Work | Blockchain type | Consensus algorithm | Trust management techniques | Blockchain storage |
|---|---|---|---|---|
| [10] | Consortium blockchain | Joint PoW/PoS consensus | Bayesian Inference Model | RSUs |
| [11] | Consortium blockchain | Joint PoW/PoS consensus | Bayesian Inference Model | RSUs |
| BTCPS [12] | Private blockchain | Joint PoW/PBFT consensus | identity-based group signature logistic regression | Authorized RSUs |
| [29] | Consortium blockchain | PBFT consensus | Anonymous cloaking region | Hyperledger |
| B-TSCA [13] | Consortium blockchain | PoW consensus | V2I authentication scheme | RSUs |
| [14] | Consortium blockchain | dPoW consensus | Physical Unclonable Functions | RSUs |
| [15] | Consortium blockchain | PBFT+PoS consensus | Mobile nodes recommendations | Fog nodes |
| VaNetChain[30] | Consortium blockchain | PoW consensus | Keccak and SHA-3 | RSUs |
| [31] | Public blockchain | Co-location consensus | Geographical area or cluster verification | Traffic servers |
| [16] | Private blockchain | PoW consensus | Direct assignment | Ethereum blockchain |
| [32] | Consortium blockchain | PoS consensus | Nearby vehicle vote | RSUs |
| BARS [33] | Public blockchain | proofs of presence | Reputation-based incentive | Vehicles+RSU |

III. BLOCKCHAIN FOR IOV PRIVACY

Privacy is one of the main requirements of IoV, the privacy of the vehicles or the information exchanged among vehicles should be preserved and revealed only to the authorized entities. While the security measures can guarantee authorized access to the vehicle's identity and exchanged messages, the challenge of privacy-preserving schemes is to maintain a balance between security level and accessibility. As shown in Figure 1, in the context of VANET, there are three main types of privacy-preserving schemes: a) identification privacy methods refer to the techniques used to prevent unauthorized entities from knowing the identity of the vehicle itself or its driver. b) location privacy methods aim to hide the location information of the vehicles from unauthorized entities, including the GPS coordinates of the current location, as well as the travelled path. c) content privacy methods aim to hide the metadata and properties of exchange messages, such as bacon messages. Many authentication schemes have been proposed for VANET. To preserve the identity of the vehicles, these schemes leverage public key infrastructure, group signature, cryptography, identity-based signature, or certificate-less signature, to name a few. These schemes depend on a centralized TA that initiates pre-set trust connection with the vehicles. The trust relationship is no longer valid when the vehicle drives to another region. To solve this issue, previous research suggested the deployment of cross-domain authentication [34], which requires various communication between vehicles, RSUs and TA, which could cause communication delays. Not to mention the additional work of protecting the centralized TA from attacks such as DoS and DDoS. The blockchain is a distributed technology that offers the possibility for decentralized privacy-preserving solutions.

Shen et al [35] introduced a blockchain privacy-preserving LBS protocol for vehicular social networks (VSN). They designed a lightweight threshold certification scheme, in which the proxy signing keys are issued by a set of CAs that act as authorized consortium blockchain nodes. Without any CA intervention, any vehicle can authenticate its identity by its blockchain address. Lu et al [36] used a blockchain-based solution to attain conditional privacy, where the vehicular nodes apply multiple certificates. The link between the real identity of the vehicle and the certificates is encrypted and securely stored in the blockchain and it can be revealed only by the law enforcement authority (LEA) in case of a dispute. Feng et al [37] introduced a blockchain-based trusted cloaking area construction (TCAC) method based on trust management to preserve the location privacy of vehicular nodes. The trust values are managed by combining vehicular regions partition, aiming to rapidly evaluate trust ratings during the cloaking area construction. George et al [38] presented a blockchain-based identity privacy framework, where the authentication parties (AP) maintain and update the blockchain. The AP generates the public and private keys along with a pseudo-ID for the vehicle and insert the transaction into the shared blockchain. Makhdoom et al. [39] proposed a content privacy solution based on blockchain, where the blockchain network is divided into multiple channels, and each channel contains a limited number of authorized entities that operate special sector. The access control rules to the data are defined by smart contracts stored in the blockchain.

TABLE II. BLOCKCHAIN FOR IOV PRIVACY

| Work | Target privacy | Privacy-preserving technique |
|---|---|---|
| [35] | Identity privacy | Threshold proxy signature scheme |
| [29] | Location privacy | Anonymous cloaking region (k-anonymity) |
| [37] | Location privacy | Trusted cloaking area construction (TCAC) |
| [39] | Content privacy | Multiple channels blockchain |
| [38] | Identity privacy | Ledger-protected pseudonyms |
| [40] | Identity privacy | Identity-based encryption |
| [41] | Content privacy | Decentralized PKI authentication |
| [42] | Identity | Distributed anonymous credentials |

| | privacy | |
|---|---|---|
| [43] | Identity privacy | Cryptographical primitives |
| [44] | Identity privacy | Conditional privacy-preserving authentication |
| [45] | Location privacy | k-anonymity privacy protection |
| [46] | Identity privacy | Group signatures and vector-based encryption |
| [47] | Identity privacy | Pseudonyms |
| [48] | Identity privacy | Pseudonyms |
| [49] | Identity privacy | Pseudonym change scheme |
| [50] | Identity privacy | Private ledger for identity information transactions |
| [51] | Identity privacy | Attribute-based blockchain |
| [52] | Identity privacy | Anonymous advertising |
| [53] | Content privacy | Flexible access control |
| [54] | Identity privacy | Anonymous vehicular announcement aggregation |
| [55] | Identity privacy | Evidence of a credible identity |
| [56] | Identity privacy | Pseudonym |
| [57] | Identity privacy | Optional traceability support |

## IV. BLOCKCHAIN FOR IoV SECURITY

Traditional vehicular network security scheme depends highly on centralized server architecture. Public key infrastructure (PKI)-based solutions need a certificate verification centralized authority, similarly, ID-based solutions required a key generation server. The former has the drawback of complicated certificate management, which is not suitable for IoV network. While the latter is prone to key escrow problems. Combining the two methods may mitigate the problem, but it is difficult to scale up hybrid solutions in practice. Blockchain was proposed as a solution to this problem. Ali et al [58] introduced Certificateless Public Key Signature (CL-PKS) which aims to reduce the computational overhead of signature verification and generation. CL-PKS leverages a bilinear pairing to offer conditional privacy-preserving authentication for V2I communication in VANETs. CL-PKS scheme also offers batch signature management and gathering signature verification functions to scale up the signature generation and verification process. Blockchain is used to execute revocation transparency of pseudo-identities prior to signature verifying. Li et al [59] combined ciphertext-based attribute encryption (CP-ABE) and block to proposed FADB, an access control scheme for VANET data based on blockchain. Specifically, they utilized user attributes to define access rights for VANET services, which improved the performance of CP-ABE. Furthermore, the lightweight vehicle attached devices can outsource heavy encryption and decryption tasks to powerful RSUs and further enhance the efficiency of the data access mechanism. Luo et al [29] introduced a blockchain-based trust management scheme using Dirichlet distribution modelling. In the proposed scheme, the vehicles establish trust connections to engage with location-based services (LBS). Vehicles send trust values to the RSUs, which execute PBFT algorithm to reach a consensus of adding a new trust block. Li et al [60] proposed a consortium blockchain-based secure authentication and key management method for VANET. Where edge computing is used for a blockchain storage system, and blockchain is utilized for V2V group key construction and real-time group membership assignment with enhanced group key updating.

## V. BLOCKCHAIN FOR IoV INCENTIVE SYSTEMS

Blockchain applications in IoV are not limited only to security and privacy. Blockchain can be used as an incentive system to reward vehicles in exchange for their participation in a given task, such as reporting traffic events, taking longer paths or choosing a non-congested electric charging station. Sankar et al [61] proposed CreditCoin, an incentive coin for VANET vehicles, which leverages a stimulus portion based on the Blockchain. Where drivers gain or invest coins when driving powers, they track legendary focuses. Furthermore, CreditCoin does jam protection and keeps the network safe by using vehicles' computational power to prevent malicious attacks. Under the same name (CreditCoin) Li et al [54] proposed an incentive announcement network based on Blockchain that aggregates vehicular announcements. CreditCoin enables different drivers to create the signatures and to send announcements anonymously in a non-trusted environment. Furthermore, CreditCoin motivates drivers with incentives to share the traffic information, which makes transactions and drivers' information tamper resistant. They have proved that CreditCoin can fulfil conditional privacy requirements and motivate users to forward announcements reliably and anonymously. Vishwakarma and Das proposed SmartCoin [62], a consortium blockchain-based incentive mechanism, where vehicles rate message source vehicles according to the validity and importance of the message. To reward useful messages, gained SmartCoins are transferred to the vehicle's account in the blockchain. The SmartCoins can be spent either at an electric charge station for electric vehicles and vehicle service stations, or a fuel station. Similarly, Khalid et al [63] introduced an incentive scheme that offers monetary incentives for vehicles that report traffic-related events. In their proposed system, RSUs verify and aggregate packets sent by vehicular nodes, and store the information related to traffic events in Interplanetary File System (IPFS). The first vehicle that initiates an event and reports gets rewarded after signature verification by RSUs. Aung et al [64] proposed T-Coin, an incentive currency used to reward drivers that opt to take longer paths to mitigate traffic congestion.

## VI. CONCLUSION

In this paper, we have reviewed recent literature on blockchain applications in the IoV. Blockchain can be leveraged to solve security challenges in IoV and ensure identification privacy as well as location privacy of the participating vehicles. Furthermore, Blockchain can also be used to manage traffic-related cryptocurrencies for rewarding vehicles that help the system. However, Blockchain-based solutions still face several challenges, such as high energy consumption in case of PoW consensus, block transitions limit, and block management overhead. The usage of blockchain in

VANET may raise various technical issues. The extremely fast movement of nodes in VANET make it difficult to establish stable connectivity, the latter is paramount requirement for some blockchain operations such as collaborative proof of stack reporting of traffic events. VANET high mobility can be tolerated by introducing a lightweight consensus mechanism that reduce transactions' validation time, but it is still challenging to achieve bazientian tolerance through consensus in limited time span.

REFERENCES

[1] S. Dhelim, H. Ning, F. Farha, L. Chen, L. Atzori, and M. Daneshmand, "IoT-Enabled Social Relationships Meet Artificial Social Intelligence," *IEEE Internet Things J.*, p. 1, 2021, doi: 10.1109/JIOT.2021.3081556.

[2] W. Zhang, N. Aung, S. Dhelim, and Y. Ai, "DIFTOS: A distributed infrastructure-free traffic optimization system based on vehicular ad hoc networks for urban environments," *Sensors (Switzerland)*, vol. 18, no. 8, 2018, doi: 10.3390/s18082567.

[3] N. Aung, W. Zhang, S. Dhelim, and Y. Ai, "Accident Prediction System Based on Hidden Markov Model for Vehicular Ad-Hoc Network in Urban Environments," *Information*, vol. 9, no. 12, p. 311, Dec. 2018, doi: 10.3390/info9120311.

[4] N. Aung, S. Dhelim, L. Chen, W. Zhang, A. Lakas, and H. Ning, "VeSoNet: Traffic-Aware Content Caching for Vehicular Social Networks based on Path Planning and Deep Reinforcement Learning," *arXiv Prepr. arXiv2111.05567*, 2021.

[5] N. Aung, W. Zhang, K. Sultan, S. Dhelim, and Y. Ai, "Dynamic traffic congestion pricing and electric vehicle charging management system for the internet of vehicles in smart cities," *Digit. Commun. Networks*, Feb. 2021, doi: 10.1016/j.dcan.2021.01.002.

[6] M. A. Bouras, Q. Lu, S. Dhelim, and H. Ning, "A Lightweight Blockchain-Based IoT Identity Management Approach," *Futur. Internet*, vol. 13, no. 2, p. 24, Jan. 2021, doi: 10.3390/fi13020024.

[7] I. Ud Din, M. Guizani, B.-S. Kim, S. Hassan, and M. Khurram Khan, "Trust Management Techniques for the Internet of Things: A Survey," *IEEE Access*, vol. 7, pp. 29763–29787, 2019, doi: 10.1109/ACCESS.2018.2880838.

[8] S. Dhelim, N. Aung, T. Kechadi, H. Ning, L. Chen, and A. Lakas, "Trust2Vec: Large-Scale IoT Trust Management System based on Signed Network Embeddings," *IEEE Internet Things J.*, pp. 1–1, 2022, doi: 10.1109/JIOT.2022.3201772.

[9] R. Hussain, J. Lee, and S. Zeadally, "Trust in VANET: A survey of current solutions and future research opportunities," *IEEE Trans. Intell. Transp. Syst.*, vol. 22, no. 5, pp. 2553–2571, 2020.

[10] H. Zhang, J. Liu, H. Zhao, P. Wang, and N. Kato, "Blockchain-based Trust Management for Internet of Vehicles," *IEEE Trans. Emerg. Top. Comput.*, pp. 1–1, 2020, doi: 10.1109/TETC.2020.3033532.

[11] Z. Yang, K. Yang, L. Lei, K. Zheng, and V. C. M. Leung, "Blockchain-based decentralized trust management in vehicular networks," *IEEE Internet Things J.*, vol. 6, no. 2, pp. 1495–1505, 2018.

[12] X. Liu, H. Huang, F. Xiao, and Z. Ma, "A Blockchain-Based Trust Management With Conditional Privacy-Preserving Announcement Scheme for VANETs," *IEEE Internet Things J.*, vol. 7, no. 5, pp. 4101–4112, May 2020, doi: 10.1109/JIOT.2019.2957421.

[13] C. Wang, J. Shen, J.-F. Lai, and J. Liu, "B-TSCA: Blockchain assisted Trustworthiness Scalable Computation for V2I Authentication in VANETs," *IEEE Trans. Emerg. Top. Comput.*, pp. 1–1, 2020, doi: 10.1109/TETC.2020.2978866.

[14] U. Javaid, M. N. Aman, and B. Sikdar, "A Scalable Protocol for Driving Trust Management in Internet of Vehicles with Blockchain," *IEEE Internet Things J.*, pp. 1–1, 2020, doi: 10.1109/JIOT.2020.3002711.

[15] D. E. Kouicem, Y. Imine, A. Bouabdallah, and H. Lakhlef, "A Decentralized Blockchain-Based Trust Management Protocol for the Internet of Things," *IEEE Trans. Dependable Secur. Comput.*, pp. 1–1, 2020, doi: 10.1109/TDSC.2020.3003232.

[16] P. K. Singh, R. Singh, S. K. Nandi, K. Z. Ghafoor, D. B. Rawat, and S. Nandi, "Blockchain-Based Adaptive Trust Management in Internet of Vehicles Using Smart Contract," *IEEE Trans. Intell. Transp. Syst.*, pp. 1–15, 2020, doi: 10.1109/TITS.2020.3004041.

[17] S. Dhelim, H. Ning, and T. Zhu, "STLF: Spatial-temporal-logical knowledge representation and object mapping framework," in *2016 IEEE International Conference on Systems, Man, and Cybernetics (SMC)*, Oct. 2016, pp. 1550–1554, doi: 10.1109/SMC.2016.7844459.

[18] S. Dhelim, H. Ning, and N. Aung, "ComPath: User Interest Mining in Heterogeneous Signed Social Networks for Internet of People," *IEEE Internet Things J.*, pp. 1–1, 2020, doi: 10.1109/JIOT.2020.3037109.

[19] S. Dhelim, H. Ning, M. A. Bouras, and J. Ma, "Cyber-Enabled Human-Centric Smart Home Architecture," in *2018 IEEE SmartWorld, Ubiquitous Intelligence & Computing, Advanced & Trusted Computing, Scalable Computing & Communications, Cloud & Big Data Computing, Internet of People and Smart City Innovations*, Oct. 2018, pp. 1880–1886.

[20] S. Dhelim, N. Aung, and H. Ning, "Mining user interest based on personality-aware hybrid filtering in social networks," *Knowledge-Based Syst.*, vol. 206, p. 106227, Oct. 2020, doi: 10.1016/j.knosys.2020.106227.

[21] S. Dhelim, H. Ning, N. Aung, R. Huang, and J. Ma, "Personality-Aware Product Recommendation System Based on User Interests Mining and Metapath Discovery," *IEEE Trans. Comput. Soc. Syst.*, vol. 8, no. 1, pp. 86–98, Feb. 2021, doi: 10.1109/TCSS.2020.3037040.

[22] S. Dhelim, N. Aung, M. A. Bouras, H. Ning, and E. Cambria, "A survey on personality-aware recommendation systems," *Artif. Intell. Rev.*, Sep. 2021, doi: 10.1007/s10462-021-10063-7.

[23] D. Wei, F. Shi, and S. Dhelim, "A Self-Supervised Learning Model for Unknown Internet Traffic Identification Based on Surge Period," *Futur. Internet*, vol. 14, no. 10, p. 289, Oct. 2022, doi: 10.3390/fi14100289.

[24] X. Cai *et al.*, "Robot and its living space A roadmap for robot development based on the view of living space," *Digit. Commun. Networks*, Dec. 2020, doi: 10.1016/j.dcan.2020.12.001.

[25] H. Ning, S. Dhelim, M. A. Bouras, A. Khelloufi, and A. Ullah, "Cyber-syndrome and its Formation, Classification, Recovery and Prevention," *IEEE Access*, 2018.

[26] S. Dhelim, N. Huansheng, S. Cui, M. Jianhua, R. Huang, and K. I.-K. Wang, "Cyberentity and its consistency in the cyber-physical-social-thinking hyperspace," *Comput. Electr. Eng.*, vol. 81, p. 106506, Jan. 2020, doi: 10.1016/j.compeleceng.2019.106506.

[27] W. Wang, H. Ning, F. Shi, S. Dhelim, W. Zhang, and L. Chen, "A Survey of Hybrid Human-Artificial Intelligence for Social Computing," *IEEE Trans. Human-Machine Syst.*, 2021.

[28] S. Dhelim, L. Chen, N. Aung, W. Zhang, and H. Ning, "A hybrid personality-aware recommendation system based on personality traits and types models," *J. Ambient Intell. Humaniz. Comput.*, pp. 1–14, Jul. 2022, doi: 10.1007/s12652-022-04200-5.

[29] B. Luo, X. Li, J. Weng, J. Guo, and J. Ma, "Blockchain Enabled Trust-Based Location Privacy Protection Scheme in VANET," *IEEE Trans. Veh. Technol.*, vol. 69, no. 2, pp. 2034–2048, Feb. 2020, doi: 10.1109/TVT.2019.2957744.

[30] P. Vintimilla-Tapia, J. Bravo-Torres, M. López-Nores, P. Gallegos-Segovia, E. Ordóñez-Morales, and M. Ramos-Cabrer, "VaNetChain: A Framework for Trustworthy Exchanges of Information in VANETs Based on Blockchain and a Virtualization Layer," *Appl. Sci.*, vol. 10, no. 21, p. 7930, Nov. 2020, doi: 10.3390/app10217930.

[31] L.-A. Hîrţan, C. Dobre, and H. González-Vélez, "Blockchain-based Reputation for Intelligent Transportation Systems," *Sensors*, vol. 20, no. 3, p. 791, Jan. 2020, doi: 10.3390/s20030791.

[32] L. Xie, Y. Ding, H. Yang, and X. Wang, "Blockchain-Based Secure and Trustworthy Internet of Things in SDN-Enabled 5G-VANETs," *IEEE Access*, vol. 7, pp. 56656–56666, 2019, doi: 10.1109/ACCESS.2019.2913682.

[33] Z. Lu, Q. Wang, G. Qu, and Z. Liu, "BARS: A Blockchain-Based Anonymous Reputation System for Trust Management in VANETs," in *2018 17th IEEE International Conference On Trust, Security And Privacy In Computing And Communications/ 12th IEEE International Conference On Big Data Science And Engineering (TrustCom/BigDataSE)*, Aug. 2018, pp. 98–103, doi: 10.1109/TrustCom/BigDataSE.2018.00025.

[34] H. Tan, S. Xuan, and I. Chung, "HCDA: Efficient Pairing-Free Homographic Key Management for Dynamic Cross-Domain Authentication in VANETs," *Symmetry (Basel).*, vol. 12, no. 6, p. 1003, 2020.

[35] H. Shen, J. Zhou, Z. Cao, X. Dong, and K.-K. R. Choo, "Blockchain-Based Lightweight Certificate Authority for Efficient Privacy-Preserving Location-Based Service in Vehicular Social Networks," *IEEE Internet Things J.*, vol. 7, no. 7, pp. 6610–6622, Jul. 2020, doi: 10.1109/JIOT.2020.2974874.

[36] Z. Lu, Q. Wang, G. Qu, H. Zhang, and Z. Liu, "A Blockchain-Based Privacy-Preserving Authentication Scheme for VANETs," *IEEE Trans. Very Large Scale Integr. Syst.*, vol. 27, no. 12, pp. 2792–2801, Dec. 2019, doi: 10.1109/TVLSI.2019.2929420.

[37] J. Feng, Y. Wang, J. Wang, and F. Ren, "Blockchain-based Data Management and Edge-assisted Trusted Cloaking Area Construction for Location Privacy Protection in Vehicular Networks," *IEEE Internet Things J.*, pp. 1–1, 2020, doi: 10.1109/JIOT.2020.3038468.

[38] S. A. George, A. Jaekel, and I. Saini, "Secure Identity Management Framework for Vehicular Ad-hoc Network using Blockchain," in *2020 IEEE Symposium on Computers and Communications (ISCC)*, Jul. 2020, pp. 1–6, doi: 10.1109/ISCC50000.2020.9219736.

[39] I. Makhdoom, I. Zhou, M. Abolhasan, J. Lipman, and W. Ni, "PrivySharing: A blockchain-based framework for privacy-preserving and secure data sharing in smart cities," *Comput. Secur.*, vol. 88, p. 101653, Jan. 2020, doi: 10.1016/j.cose.2019.101653.

[40] Y. Pu, T. Xiang, C. Hu, A. Alrawais, and H. Yan, "An efficient blockchain-based privacy preserving scheme for vehicular social networks," *Inf. Sci. (Ny).*, vol. 540, pp. 308–324, Nov. 2020, doi: 10.1016/j.ins.2020.05.087.

[41] Y. Li and B. Hu, "A Consortium Blockchain-enabled Secure and Privacy-Preserving Optimized Charging and Discharging Trading Scheme for Electric Vehicles," *IEEE Trans. Ind. Informatics*, pp. 1–1, 2020, doi: 10.1109/TII.2020.2990732.

[42] L. Wang, X. Lin, E. Zima, and C. Ma, "Towards Airbnb-Like Privacy-Enhanced Private Parking Spot Sharing Based on Blockchain," *IEEE Trans. Veh. Technol.*, vol. 69, no. 3, pp. 2411–2423, Mar. 2020, doi: 10.1109/TVT.2020.2964526.

[43] K. Shi, L. Zhu, C. Zhang, L. Xu, and F. Gao, "Blockchain-based multimedia sharing in vehicular social networks with privacy protection," *Multimed. Tools Appl.*, vol. 79, no. 11–12, pp. 8085–8105, Mar. 2020, doi: 10.1007/s11042-019-08284-8.

[44] C. Lin, D. He, X. Huang, N. Kumar, and K.-K. R. Choo, "BCPPA: A Blockchain-Based Conditional Privacy-Preserving Authentication Protocol for Vehicular Ad Hoc Networks," *IEEE Trans. Intell. Transp. Syst.*, pp. 1–13, 2020, doi: 10.1109/TITS.2020.3002096.

[45] Y. Qiu, Y. Liu, X. Li, and J. Chen, "A Novel Location Privacy-Preserving Approach Based on Blockchain," *Sensors*, vol. 20, no. 12, p. 3519, Jun. 2020, doi: 10.3390/s20123519.

[46] C. Zhang et al., "BSFP: Blockchain-Enabled Smart Parking With Fairness, Reliability and Privacy Protection," *IEEE Trans. Veh. Technol.*, vol. 69, no. 6, pp. 6578–6591, Jun. 2020, doi: 10.1109/TVT.2020.2984621.

[47] M. U. Javed, A. Jamal, N. Javaid, N. Haider, and M. Imran, "Conditional Anonymity enabled Blockchain-based Ad Dissemination in Vehicular Ad-hoc Network," in *2020 International Wireless Communications and Mobile Computing (IWCMC)*, Jun. 2020, pp. 2149–2153, doi: 10.1109/IWCMC48107.2020.9148487.

[48] D. Zheng, C. Jing, R. Guo, S. Gao, and L. Wang, "A Traceable Blockchain-Based Access Authentication System With Privacy Preservation in VANETs," *IEEE Access*, vol. 7, pp. 117716–117726, 2019, doi: 10.1109/ACCESS.2019.2936575.

[49] S. Bao et al., "Pseudonym Management Through Blockchain: Cost-Efficient Privacy Preservation on Intelligent Transportation Systems," *IEEE Access*, vol. 7, pp. 80390–80403, 2019, doi: 10.1109/ACCESS.2019.2921605.

[50] B. Guehguih and H. Lu, "Blockchain-Based Privacy-Preserving Authentication and Message Dissemination Scheme for VANET," in *Proceedings of the 2019 5th International Conference on Systems, Control and Communications*, Dec. 2019, pp. 16–21, doi: 10.1145/3377458.3377466.

[51] L. Cheng et al., "SCTSC: A Semicentralized Traffic Signal Control Mode With Attribute-Based Blockchain in IoVs," *IEEE Trans. Comput. Soc. Syst.*, vol. 6, no. 6, pp. 1373–1385, Dec. 2019, doi: 10.1109/TCSS.2019.2904633.

[52] W. Yang, X. Dai, J. Xiao, and H. Jin, "LDV: A Lightweight DAG-Based Blockchain for Vehicular Social Networks," *IEEE Trans. Veh. Technol.*, vol. 69, no. 6, pp. 5749–5759, Jun. 2020, doi: 10.1109/TVT.2020.2963906.

[53] X. Ma, C. Ge, and Z. Liu, "Blockchain-Enabled Privacy-Preserving Internet of Vehicles: Decentralized and Reputation-Based Network Architecture," in *International Conference on Network and System Security*, 2019, pp. 336–351.

[54] L. Li et al., "CreditCoin: A Privacy-Preserving Blockchain-Based Incentive Announcement Network for Communications of Smart Vehicles," *IEEE Trans. Intell. Transp. Syst.*, vol. 19, no. 7, pp. 2204–2220, Jul. 2018, doi: 10.1109/TITS.2017.2777990.

[55] C. Xu, H. Liu, P. Li, and P. Wang, "A Remote Attestation Security Model Based on Privacy-Preserving Blockchain for V2X," *IEEE Access*, vol. 6, pp. 67809–67818, 2018, doi: 10.1109/ACCESS.2018.2878995.

[56] N. Malik, P. Nanda, A. Arora, X. He, and D. Puthal, "Blockchain Based Secured Identity Authentication and Expeditious Revocation Framework for Vehicular Networks," in *2018 17th IEEE International Conference On Trust, Security And Privacy In Computing And Communications/ 12th IEEE International Conference On Big Data Science And Engineering (TrustCom/BigDataSE)*, Aug. 2018, pp. 674–679, doi: 10.1109/TrustCom/BigDataSE.2018.00099.

[57] R. Sharma and S. Chakraborty, "BlockAPP: Using Blockchain for Authentication and Privacy Preservation in IoV," in *2018 IEEE Globecom Workshops (GC Wkshps)*, Dec. 2018, pp. 1–6, doi: 10.1109/GLOCOMW.2018.8644389.

[58] I. Ali, M. Gervais, E. Ahene, and F. Li, "A blockchain-based certificateless public key signature scheme for vehicle-to-infrastructure communication in VANETs," *J. Syst. Archit.*, vol. 99, p. 101636, 2019.

[59] H. Li, L. Pei, D. Liao, S. Chen, M. Zhang, and D. Xu, "FADB: A fine-grained access control scheme for VANET data based on blockchain," *IEEE Access*, vol. 8, pp. 85190–85203, 2020.

[60] X. Li and X. Yin, "Blockchain-based group key agreement protocol for vehicular ad hoc networks," *Comput. Commun.*, vol. 183, pp. 107–120, 2022.

[61] P. Phani Sankar, P. Anil Kumar, and B. Bharathi, "Blockchain-Based Incentive Announcement In Vanet Using CreditCoin," in *Advances in Electronics, Communication and Computing*, Springer, 2021, pp. 567–574.

[62] L. Vishwakarma and D. Das, "SmartCoin: A novel incentive mechanism for vehicles in intelligent transportation system based on consortium blockchain," *Veh. Commun.*, vol. 33, p. 100429, 2022.

[63] A. Khalid, M. S. Iftikhar, A. Almogren, R. Khalid, M. K. Afzal, and N. Javaid, "A blockchain based incentive provisioning scheme for traffic event validation and information storage in VANETs," *Inf. Process. \& Manag.*, vol. 58, no. 2, p. 102464, 2021.

[64] N. Aung, W. Zhang, S. Dhelim, and Y. Ai, "T-Coin: Dynamic Traffic Congestion Pricing System for the Internet of Vehicles in Smart Cities," *Information*, vol. 11, no. 3, p. 149, Mar. 2020, doi: 10.3390/info11030149.